\documentclass[]{rmf-d}
\usepackage{nopageno,rmfbib,multicol,times,epsf,amsmath,amssymb} 

\usepackage[]{caption2}
\usepackage{graphics}

\def\ve{\varepsilon}

\def\et{\eta}

\def\rh{\rho}

\def\si{\sigma}

\def\om{\omega}

\def\cl{{\cal L}}

\def\fr#1#2{{{#1} \over {#2}}}
\def\half{{\textstyle{1\over 2}}}

\def\lsim{\mathrel{\rlap{\lower3pt\hbox{$\sim$}}
    \raise2pt\hbox{$<$}}}
\def\gsim{\mathrel{\rlap{\lower3pt\hbox{$\sim$}}
    \raise2pt\hbox{$>$}}}
\def\sqr#1#2{{\vcenter{\vbox{\hrule height.#2pt
         \hbox{\vrule width.#2pt height#1pt \kern#1pt
         \vrule width.#2pt}
         \hrule height.#2pt}}}}

\def\prt{\partial}
\def\lrpartial{\raise 1pt\hbox{$\stackrel\leftrightarrow\partial$}}

\def\Re{\hbox{Re}\,}
\def\Im{\hbox{Im}\,}

\def\etal{{\it et al.}}

\newcommand{\bit}{\begin{itemize}}
\newcommand{\eit}{\end{itemize}}

\newcommand{\beq}[1]{\begin{equation}\label{#1}}
\newcommand{\eeq}{\end{equation}}
\newcommand{\bea}[1]{\begin{eqnarray}\label{#1}}
\newcommand{\eea}{\end{eqnarray}}
\newcommand{\ba}{\begin{array}}
\newcommand{\ea}{\end{array}}

\newcommand{\rf}[1]{(\ref{#1})}

\def\fr#1#2{{{#1} \over {#2}}}
\def\etal{{\it et al.}}

\def\rmfcornisa{INVESTIGACI\'ON \hfill\rmf\ {\bf 56} (6) 469--474 

\hfill Diciembre 2010}

\def\rmfcintilla{{\it Rev.\ Mex.\ F\'{\i}s.\/} {\bf 56} (6) (2010) 469--474}
\clearpage 
\rmfcaptionstyle 
\begin{document}

\title{A mapping between Lorentz-violating and conventional electrodynamics 
\vspace{-6pt}}

\author{Ralf Lehnert}
\address{Instituto de Ciencias Nucleares\\
Universidad Nacional Aut\'onoma de M\'exico\\
Apartado Postal 70-543, M\'exico, 04510, D.F., Mexico}

\maketitle \recibido{el 2 de agosto de 2010}{7 de septiembre de 2010\vspace{-12pt}}

\begin{abstract}
The Chern--Simons-type term in the photon sector 
of the Lorentz- and CPT-breaking minimal Standard-Model Extension (mSME)
is considered.
It is argued that 
under certain circumstances 
this term can be removed from the mSME.
In particular, 
it is demonstrated that for lightlike Lorentz violation
a field redefinition exists 
that maps the on-shell free Chern--Simons model 
to conventional on-shell free electrodynamics. 
A compact explicit expression for an operator implementing such a mapping 
is constructed.
This expression establishes 
that the field redefinition is non-local. 
\end{abstract}
\keys{Lorentz violation; CPT violation; Standard-Model Extension; field redefinition. \vspace{-4pt}}
\begin{resumen}
Se considera el t\'ermino de tipo Chern--Simons 
en el sector fot\'onico de la extensi\'on m\'{\i}nima 
del modelo est\'andar con rompimiento de Lorentz y de CPT. 
Se discuten las circunstancias bajo las cuales este t\'ermino 
se puede remover del modelo. 
En particular, 
se demuestra que para 
una violaci\'on de Lorentz parametrizada por un cuadrivector tipo luz
existe una redefinici\'on de campo 
que proyecta el modelo de Chern--Simons 
libre en la capa de masas 
a la electrodin\'amica 
convencional 
libre en la capa de masas.
Se construye una expres\'on compacta expl\'{\i}cita 
para un operador implementando tal proyecci\'on. 
Esta expresi\'on establece que la redefinici\'on de campo es no-local.
\end{resumen}
\descript{Violaci\'on de Lorentz; violaci\'on CPT; Modelo Est\'andar Extendido; redefinici\'on de campo. \vspace{-4pt}}
\pacs{11.30.Cp, 11.30.Er, 12.60.-i \vspace{-4pt}}
\begin{multicols}{2}

\section{Introduction}
\label{intro}

Despite its phenomenological successes, 
the present framework for fundamental physics---the Standard Model of particle 
physics together with the General Theory of Relativity---leaves unanswered 
various conceptual questions. 
For this reason, 
a substantial amount of theoretical work 
is currently being devoted to the search for an underlying theory 
that provides a quantum description of gravity. 
Experimental tests of such ideas face, however, 
a considerable obstacle of practical nature: 
most quantum-gravity predictions in virtually every leading candidate model 
are expected to be extremely small due to the anticipated Planck-scale suppression. 

During the last two decades, 
a variety of theoretical investigations 
have suggested the possibility of spacetime-symmetry breakdown 
in leading candidate models for underlying physics.
Examples of such investigations involve  
string field theory\cite{ksp}, 
realistic field theories on noncommutative spacetimes\cite{ncqed}, 
cosmologically varying scalars\cite{spacetimevarying}, 
various quantum-gravity models\cite{qg}, 
four-dimensional spacetimes with a nontrivial topology\cite{klink}, 
random-dynamics models\cite{fn02}, 
multiverses\cite{bj}, 
and brane-world scenarios\cite{brane}. 
Although the dynamical structures underlying the above models 
typically remain Lorentz symmetric, 
Lorentz and CPT violation can nevertheless occur 
in the ground state at low energies. 
These ideas 
provide a key motivation 
for Lorentz- and CPT- violation searches 
in the context of quantum gravity.

At energies 
that can currently be reached in experimental situations, 
the effects resulting from Lorentz and CPT violation 
in underlying physics
can be described by the Standard-Model Extension (SME), 
which is an 
effective-field-theory framework 
containing the usual Standard Model\cite{flatsme} 
and General Relativity\cite{curvedsme} 
as limiting cases. 
The minimal SME (mSME), 
which only contains relevant and marginal operators, 
has provided the basis for
numerous experimental investigations of Lorentz- and CPT-symmetry\cite{review}. 
Specific studies include, 
for instance, 
ones with photons\cite{randomphotonexpt,cherenkov}, 
neutrinos\cite{randomnuexpt},
electrons\cite{randomeexpt}, 
protons and neutrons\cite{randompnexpt}, 
mesons\cite{randomhadronexpt}, 
muons\cite{muexpt},
and gravity\cite{gravity}.  
Several of the obtained experimental limits 
can be regarded as testing Planck-scale physics.

Internal consistency and a thorough theoretical understanding 
are of key importance for test models, 
such as the SME.
For this reason,
a number of SME investigations 
have addressed such questions\cite{investigations,rl06,hariton07}. 
Some of these studies 
have shed light on various conceptual questions,  
but so far none have
suggested any internal inconsistencies. 
It nevertheless remains necessary 
to keep illuminating the internal structure of the SME, 
both to gain insight into Lorentz and CPT violation 
and to solidify further the theoretical basis of the SME.
In this context, 
the theory of free particles 
is of particular interest:
They correspond to external legs in Feynman diagrams 
and are therefore an important theoretical ingredient in perturbative QFT. 
Moreover, 
many experimental studies, 
such as kinematical cosmic-ray tests of Lorentz symmetry, 
rely to a large part on free-particle physics. 
This work aims at illuminating
various aspects of the Chern--Simons-type term 
contained in the free-photon sector of the mSME. 
More specifically, 
we will show that this term can be removed on-shell 
from this sector under certain conditions.

The outline of this work is as follows.
The Chern--Simons limit of the mSME is briefly reviewed in Sec.~2. 
Section~3 argues that for lightlike Lorentz violation,
the Chern--Simons term can be removed on-shell 
from the free model, 
and the idea behind the associated mapping is illustrated. 
Section~4 derives a compact expression 
for this field-redefinition mapping. 
A summary and a brief outlook are contained in Sec.~5.

\section{Basics}
\label{model}

A particularly popular limit of the mSME 
is the Lorentz- and CPT-violating Chern--Simons extension
of electrodynamics. 
This limit will be the focus 
of the subsequent discussion in this work,  
so we begin by 
reviewing various known results pertaining to Chern--Simons electrodynamics.
Adopting natural units 
$c\hspace{-1pt} =\hspace{-1pt} \hbar\hspace{-1pt} =\hspace{-1pt} 1$
and the metric signature $(+,-,-,-)$,
the free model Lagrangian
is given by \cite{mcs}
\beq{lagr}
\cl_{\rm MCS} = 
-\fr{1}{4} F^2
+(k_{AF})^{\mu}A^{\nu}\tilde{F}_{\mu\nu}\,,
\eeq
where $F_{\mu\nu}=\prt_{\mu}A_{\nu}-\prt_{\nu}A_{\mu}$ 
and $\tilde{F}^{\mu\nu}=\half\ve^{\mu\nu\rh\si}F_{\rh\si}$, 
as usual.
The nondynamical fixed $(k_{AF})^{\mu}$ 
selects a preferred direction in spacetime 
explicitly breaking Lorentz and CPT symmetry. 
In what follows, 
we will drop the $AF$ subscript 
for brevity.
Although this Lagrangian is gauge dependent, 
the corresponding action integral, 
and therefore the physics,
are invariant
under gauge transformations.

The Euler--Lagrange equations
associated with the Lagrangian \rf{lagr} 
yield the following
equations of motion for the potentials $A^{\mu}=(A^0,\vec{A})$: 
\beq{oddeom}
\left(\Box \et^{\mu\nu}-\prt^{\mu} \prt^{\nu}
-2\,\ve^{\mu\nu\rh\si}k_{\rh}\prt_{\si}\right)A_{\nu}
=0\, .
\eeq
From Eq.~\rf{oddeom}, 
the Chern--Simons modified Maxwell equations 
\beq{maxwell}
\partial_{\mu}F^{\mu\nu}+2\,k_{\mu}\tilde{F}^{\mu\nu}=0 
\eeq
can be derived, 
which put into evidence 
the gauge invariance of the model. 
For completeness, 
we also exhibit the modified Coulomb 
and Amp\`ere laws, 
which are contained in Eq.\ \rf{maxwell}: 
\bea{oddmax}
\vec{\nabla}\!\cdot\!\vec{E}-2\vec{k}\!\hspace{0.8pt}\cdot\!\vec{B} & = & 0\, ,\nonumber\\
-\dot{\hspace{-1pt}\vec{E}}+\vec{\nabla}\!\times\!\vec{B}
-2k_0\vec{B}+2\vec{k}\!\times\!\vec{E} & = & 
0\, .
\eea
The homogeneous Maxwell equations 
are left unchanged 
because the relationship between the fields and potentials is the conventional one. 

Paralleling the ordinary Maxwell case, 
$A^0$ is nondynamical, 
and gauge symmetry eliminates another mode of $A^{\mu}$, 
so that Eq.\ \rf{oddeom} contains only two independent degrees of freedom.
It is often convenient to fix a gauge, 
and we will actually do so in the next section.
It turns out 
that any of the usual conditions on $A^{\mu}$, 
such as Lorentz or Coulomb gauge,  
can be imposed.
We remark, 
however, 
that there are some differences 
between conventional electrodynamics and the Chern--Simons model 
regarding the equivalence of certain gauge choices.
A more detailed discussion 
of the degrees of freedom and the gauge-fixing process 
can be found in the second paper of Ref.\cite{flatsme}.

The plane-wave dispersion relation can be obtained
with the ansatz $A^{\mu}(x)=\varepsilon^{\mu}(\lambda)\exp (-i\lambda\! \cdot\! x)$, 
where $\lambda^{\mu}\equiv(\omega,\vec{\lambda})$. 
This ansatz and the equations of motion \rf{oddeom} 
give
\beq{odddisp}
\lambda^4+4\lambda^2k^2-4(\lambda\!\hspace{0.8pt}\cdot\! k)^2=0\, .
\eeq
This equation determines the wave frequency $\om$
as a function of the wave 3-vector $\vec{\lambda}$.

Examples of non-standard effects 
caused by the inclusion of the Chern--Simons term into electrodynamics
are vacuum birefringence\cite{mcs} 
and vacuum Cherenkov radiation\cite{cherenkov}.
We also mention 
that for a timelike $k^{\mu}$, 
the magnitude of the group velocity 
determined by the dispersion relation \rf{odddisp} 
can in certain circumstances exceed the light speed $c$. 
This is consistent with previous analyses\cite{mcs,adamklink01},
which have established theoretical difficulties 
associated with instabilities and causality violations 
for $k^2>0$. 
Such issues do not arise for $k^2\leq 0$. 
In what follows, 
we focus primarily on the case of a lightlike $k^{\mu}$. 
We remark in passing 
that the lightlike case 
possesses more residual spacetime symmetries 
than the timelike and spacelike cases\cite{hariton07}.

\section{General idea behind the mapping}
\label{idea} 

Our goal  
is to establish 
that a lightlike $k^{\mu}$ 
can be removed from the (free) equations of motions \rf{oddeom} 
by an on-shell field redefinition. 
This section discusses certain features of the solutions to Eq.\ \rf{oddeom} 
that give some intuition 
as to why $k^{\mu}$ is removable 
thereby motivating the form of the field redefinition. 

For a lightlike $k^{\mu}$, 
the dispersion relation \rf{odddisp} 
can be cast into the form 
\beq{lightlikedr}
\big[\lambda^{\mu}+(-1)^a k^{\mu}\big]^2=0\;,
\eeq
where $a=1,2$. 
This equation possesses the roots 
\beq{roots}
(\lambda^\pm_a)^{\mu}=\Big((\lambda^\pm_a)^0(\vec{\lambda}),\vec{\lambda}\Big)
\eeq
with 
\beq{dispsoln} 
(\lambda^\pm_a)^0(\vec{\lambda})=\pm\big| \, \vec{\lambda}\pm(-1)^a\vec{k} \, \big|\mp(-1)^ak^0\;.
\eeq 
Here, $k^\mu=(k^0,\vec{k})$ is lightlike $|k^0|=|\vec{k}|$, 
and we have chosen a more convenient labeling of these solutions
than that given in Eq.~\rf{lightlikedr}.
The roots $(\lambda^\pm_a)^{\mu}$ may alternatively be parametrized as
\beq{dispsoln2}
(\lambda^\pm_a)^{\mu}=p_\pm^{\mu}\mp(-1)^a k^{\mu}\;.
\eeq
Here, $p_\pm^{\mu}\equiv(\pm|\vec{p}\,|,\vec{p})$ satisfies the conventional photon dispersion relation $(p_\pm)^2=0$, 
where the subscripts $+$ and $-$ label the positive- and negative-frequency solutions, respectively. 
This result is immediately evident from Eq.\ \rf{lightlikedr} 
when the expression inside the square brackets is identified 
with the appropriate $p_\pm^{\mu}$. 

Equation~\rf{dispsoln} determines four roots 
labeled by $a=1,2$ and the subscript $\pm$, 
which are seemingly independent.
This reflects the fact that 
the dispersion relation~\rf{odddisp} is quartic in $\lambda^0$.
One might then wonder 
whether this is consistent 
with the previous statement in Sec.~2 
that our Chern--Simons model contains two independent degrees of freedom, 
just like conventional electrodynamics.
To resolve this apparent contradiction, 
the two negative-frequency roots 
must be properly interpreted.
To this end, 
recall that in QFT the negative-energy solutions
are regarded as positive-energy reversed-momentum states 
corresponding to antiparticles.
Defining $(\lambda^\pm_a)^0(\vec{\lambda})\equiv\pm\omega_a(\pm\vec{q})$ 
leaves unaffected the positive-frequency roots in Eq.~\rf{dispsoln}
\beq{posfreqroots}
\omega_a(\vec{q})=\big| \, \vec{q}+(-1)^a\vec{k} \, \big|-(-1)^ak^0\;.
\eeq
The negative-valued roots given by the lower sign in Eq.~\rf{dispsoln}
take the form
\beq{negfreqroots}
-\omega_a(\vec{q})=-\big| \, -\vec{q}-(-1)^a\vec{k} \, \big|+(-1)^ak^0\;.
\eeq
Inspection shows that
with this conventional reinterpretation
Eqs.~\rf{posfreqroots} and~\rf{negfreqroots} 
are identical.
Moreover, 
there are only two independent polarization vectors
($A^0$ is non-dynamical, 
and a choice of gauge places an additional constraint on $A^\mu$).
This establishes that 
the physics described by the two negative-energy solutions 
must be identical to the physics contained in the two positive-energy solutions, 
as expected.
In what follows, 
we may therefore focus solely 
on the positive-frequency solutions 
and omit the subscript $\pm$ from now on. 
We remark that $a=1,2$ 
labels the two helicity-type polarizations states 
of plane waves. 

The above reparametrization~\rf{dispsoln2}
may then be taken to read $(\lambda_a)^{\mu}=p^{\mu}-(-1)^a k^{\mu}$, 
where $p^\mu\equiv p^\mu_+$ has a positive-valued zeroth component 
and continues to satisfy $p^\mu p_\mu=0$.
This reparametrization is the key 
to an intuitive understanding of the field redefinition
that removes $k^{\mu}$ from the equations of motion: 
up to polarization vectors, 
any plane-wave exponential that solves Eq.~\rf{oddeom} 
is of the form
\beq{pwexp}
\exp(-i\lambda_a\!\cdot\!x)=\exp(-ip\!\cdot\!x)\exp[+(-1)^a ik\!\cdot\!x]\,. 
\eeq
Note $\exp(-ip\!\cdot\! x)$ corresponds to Lorentz-symmetric massless plane waves;
the Lorentz-violating contribution $\exp[+(-1)^a ik\!\cdot\!x]$ 
can be removed via a field redefinition
\beq{schematic}
(\textrm{plane wave})\;\to\;(\textrm{plane wave})\,\exp\left[-(-1)^a ik\!\cdot\!x\right]\,.
\eeq
We remark 
that this field redefinition depends only on the plane-wave label $a$;
it is independent of the plane-wave momentum. 
In other words, 
any superpositions of plane-wave exponentials 
with label $a=1$
can be redefined by removing the common $\exp(-ik \cdot x)$ factor, 
and superpositions of plane-wave exponentials 
with labels $a=2$
can be redefined by removing the common $\exp(+ik\!\cdot\! x)$ factor. 

To make this idea more precise, 
consider the general explicit solution to the free equations of motions: 
\beq{gensoln} 
A^{\mu}(x)=\int\!\frac{d^3 \vec{\lambda}}{(2\pi)^3}
\!\!\sum_{a=1,2}\big[\epsilon_{a}^{\mu}(\vec{\lambda})e^{-i\lambda_a\cdot x}
+{\epsilon_{a}^{\mu}}^*(\vec{\lambda})e^{+i\lambda_a \cdot x}\big]\,. 
\eeq 
The polarization vectors $\epsilon_{a}^{\mu}(\vec{\lambda})$ 
are constrained by the equations of motion, 
the gauge choice, 
and---in cases of degeneracy---by the requirement of linear independence. 
We have absorbed the relativistic normalization factor 
of the integration measure 
into the definition of the $\epsilon_{a}^{\mu}(\vec{\lambda})$, 
so that they do not transform as 4-vectors. 
In what follows, 
we will nevertheless continue to refer to these quantities as polarization vectors.

With Eq.~\rf{dispsoln2} at hand, 
we may change integration variables from $\vec{\lambda}$ to $\vec{p}$
in Eq.\ \rf{gensoln}.
Note that this is 
just a linear shift, 
so that the Jacobian is trivial. 
The exponentials will now contain a $\vec{p}$-independent piece, 
which can be pulled out of the integral: 
\beq{gensoln2} 
A^{\mu}(x)={\cal A}^{\mu}(x)\exp(-ik\!\cdot\!x)+{{\cal A}^{\mu}}^*(x)\exp(+ik\!\cdot\!x)\;,
\eeq
where 
\beq{Aadef} 
{\cal A}^{\mu}(x)\equiv\int\!\frac{d^3 p}{(2\pi)^3}
\left[\xi_{1}^{\mu}(\vec{p})e^{-ip\cdot x}
+{\xi_{2}^{\mu}}^*(\vec{p})e^{+ip\cdot x}\right]. 
\eeq 
In this expression, 
the new polarization vectors $\xi_{a}^{\mu}(\vec{p})$
are given in terms of the old polarization vectors $\epsilon_{a}^{\mu}(\vec{\lambda})$ 
simply by a shift in the argument 
$\xi_{a}^{\mu}(\vec{p})=\epsilon_{a}^{\mu}(\vec{p}-[-1]^a \vec{k})$. 

The Definition \rf{Aadef} reveals 
that the fields ${\cal A}^{\mu}(x)$ 
are superpositions of plane waves 
with Lorentz-symmetric dispersion relation $p^\mu p_\mu=0$. 
This fact implies that 
$\Box {\cal A}^{\mu}(x)=0$. 
Note 
that this equation 
resembles the conventional Maxwell equations in Lorentz gauge. 
As advertised above, 
this field is obtained from the original solution $A^{\mu}$ 
by first splitting off all exponentials with the common factor $\exp(-ik\!\cdot\!x)$ 
to find ${\cal A}^{\mu}(x)$ and subsequently removing this factor.
An analogous procedure must be performed for ${{\cal A}^{\mu}}^*(x)$.

The complex-valued ${\cal A}^{\mu}(x)$ field 
gives rise to a real-valued vector field $\underline{A}^{\mu}(x)$ 
in a natural way:
\bea{redefA}
 \! \! \! \! \! \! \! \! \! \! \! 
 \underline{A}^{\mu}(x) & \! \! \! \!  = \! \! \! \!  & 
 {\cal A}^{\mu}(x)+{{\cal A}^{\mu}}^*(x)\nonumber\\
& \! \! \! \! \! = \! \! \! \! \! &  \! \int \! \! \frac{d^3 \vec{p}}{(2\pi)^3}
 \! \! \sum_{a=1,2} \! \!\left[\xi_{a}^{\mu}(\vec{p})e^{-ip \cdot  x}\!
+{\xi_{a}^{\mu}}^*\!(\vec{p})e^{+ip \cdot  x}\right].
\eea
This field also obeys an equation 
that is consistent with the conventional Maxwell theory in Lorentz gauge: 
\beq{eomA}
\Box \underline{A}^{\mu}(x)=0\,.
\eeq
We therefore see 
that a given solution $A^{\mu}(x)$ of the Chern--Simons modified electrodynamics 
leads to a field $\underline{A}^{\mu}(x)$ 
that obeys a Klein--Gordon-type equation in each component, 
so at least the plane-wave exponentials are Lorentz symmetric.

Equation~\rf{eomA} essentially governs the spacetime dependence  
of the redefined field $\underline{A}^\mu(x)$ via the plane-wave exponentials, 
but it leaves undetermined the polarizations vectors. 
This is consistent with the gauge invariance of the Chern--Simons model: 
we have not yet selected a gauge for $A^{\mu}$, 
but the field redefinition gives a field $\underline{A}^\mu$ 
satisfying Eq.~\rf{eomA}, 
which looks gauge fixed (Lorentz gauge). 
This is, 
of course, 
not the case 
precisely because of the above issue 
that the polarizations vectors are still undetermined. 
For $\underline{A}^\mu$ to obey the usual Maxwell equations in Lorentz gauge, 
we not only need Eq.~\rf{eomA}, 
but also the additional Lorentz condition $\partial_\mu\underline{A}^\mu=0$. 
This condition constrains the polarization vectors $\xi_{a}^{\mu}(\vec{p})$ 
to be transverse. 

Suppose we select Lorentz gauge in the Chern--Simons model $\partial_\mu A^\mu=0$. 
Then,
the question arises 
as to whether our field redefinition leaves unchanged this gauge condition. 
This is, 
in fact, 
not the case. 
We obtain 
\beq{Lorentzgaugemap}
\partial_\mu A^\mu=0\quad\to\quad\partial_\mu\underline{A}^\mu=-2\, \Im (k\!\cdot\!{\cal A})
\eeq
for the redefined condition. 
Since $\Im ({\cal A}^{\mu})$ cannot be freely chosen 
(it is determined by $\Re({\cal A}^{\mu})$ to yield plane-wave exponentials), 
the redefined field $\underline{A}^\mu$ fails to obey the Lorentz condition. 
Let us instead select the gauge 
\beq{othergauge}
\partial_\mu A^\mu(x)=2\, \Im\! \left[\,k\!\cdot\!{\cal A}(x)\,\exp(-ik\!\cdot\!x)\,\right]
\eeq
for the solution of the Chern--Simons model. 
Substituting Eq.~\rf{gensoln2} on the left-hand side of Eq.~\rf{othergauge} 
then gives $\Re\!\left[\,\exp(-ik\!\cdot\!x)\,\partial\!\cdot\!{\cal A}(x)\,\right]=0$.
Using the plane-wave expansion of ${\cal A}^\mu (x)$,  
one can verify\footnote{We exclude special cases, 
such as solutions containing a single plane wave with $p^{\mu}=\pm k^{\mu}$ and specially chosen phases 
or solutions of two plane waves with $p^{\mu}=0$ and specially chosen phases.} 
that this essentially implies 
$\partial\!\cdot\!{\cal A}(x)=\partial\!\cdot\!{\cal A}^*(x)=0$,
and therefore
\beq{othergaugeimplication}
\partial_\mu \underline{A}^\mu(x)=0\,.
\eeq
This result establishes 
that with a carefully selected gauge for solutions $A^{\mu}$ in the Chern--Simons model, 
the field redefinition discussed above 
yields a solution $\underline{A}^{\mu}$ of conventional electrodynamics in Lorentz gauge. 
It follows 
that such a mapping, defined on-shell, 
removes Lorentz and CPT violation from the Chern--Simons model.

\section{Compact expression for the mapping}
\label{expression} 

In the previous section, 
we have discussed the possibility of 
removing a lightlike Lorentz- and CPT-violating $k^\mu$  
from Chern--Simons electrodynamics. 
We have illustrated why and how this can be achieved. 
The basic idea has been the following. 
The first step is to decompose an arbitrary solution $A^\mu$ of the Chern--Simons model 
into two pieces, 
one containing the plane waves with dispersion relation shifted by $+k^\mu$
and the other containing those with the opposite shift $-k^\mu$. 
In the second step, 
the Lorentz-violating shift 
is undone with a simple multiplicative field redefinition involving $\exp(+ik\!\cdot\!x)$ 
in one of these pieces and $\exp(-ik\!\cdot\!x)$ in the other.

We now set out to find a more compact form for such an on-shell mapping 
from the set of solutions in the Chern--Simons model 
to the set of solutions in ordinary electrodynamics. 
Clearly, 
the more challenging step in the field redefinition 
is the first one, 
which decomposes a given solution $A^\mu$ 
according to the shift direction in $k^\mu$ 
yielding both ${\cal A}^\mu$ and ${{\cal A}^\mu}^*$. 
In principle, 
this task can be performed with Fourier methods 
that simply project out the desired pieces. 
This section gives closed-form expressions for suitable projectors. 

We begin by characterizing the set of solutions $A^{\mu}$ to Eq.~\rf{oddeom}, 
which we take as the domain for our field-redefinition mapping. 
As before, 
we fix $k^\mu$ to be lightlike $k^\mu k_\mu=0$, 
and we consider all well-behaved fields 
of the form displayed in Eq.~\rf{gensoln}. 
Note that all plane-wave momenta in the exponentials 
satisfy the dispersion relation~\rf{lightlikedr}. 
In particular, 
any field $A^{\mu}$ obeying Eq.~\rf{gensoln} 
therefore also satisfies
\beq{lightlikeeom}
\left[\Box^2+4(k\!\cdot\!\partial)^2
\right]A^\mu(x)=0\,.
\eeq
We remark 
that all other solutions to Eq.~\rf{oddeom} 
can differ from Eq.~\rf{gensoln}
only by a total derivative, 
a quantity that leaves unaffected the physics. 
Note 
that we are not committing ourselves 
to a definite gauge 
because of the remaining freedom in the choice of polarization vectors in Eq.~\rf{gensoln}.

Consider the operators $P_+$ and $P_-$  defined 
by
\beq{Pdef}
P_\pm\equiv\fr{1}{2}\left(1\pm2i\,\fr{k\!\cdot\!\partial}{\Box}\,\right)\,.
\eeq
When these operators act on our set of solution $A^\mu$ 
characterized above, 
we find the following:
the operators are complete in the sense that $P_++P_-=1$,
they are orthogonal in the sense $P_\pm P_\mp=0$, 
and they are idempotent $P_\pm^2=P_\pm$. 
To arrive at these results, 
we have employed Eq.~\rf{lightlikeeom}. 
It is apparent 
that $P_+$ and $P_-$ are operators 
that project onto two disjoint subsets 
of solutions. 
Moreover, 
the union of these subsets 
is equal to our full set of solutions. 

Applying $P_\pm$ to the plane-wave exponentials 
occurring in the general solution~\rf{gensoln}
yields
\bea{expaction}
P_+\,e^{\pm i \lambda_a \cdot x} & = & \frac{1}{2}\left[1\pm(-1)^a\right]e^{\pm i \lambda_a \cdot x}\,,\nonumber\\
P_-\,e^{\pm i \lambda_a \cdot x} & = & \frac{1}{2}\left[1\mp(-1)^a\right]e^{\pm i \lambda_a \cdot x}\,.
\eea
Employing these relations, 
one can then show that 
\bea{aaction}
P_+\,A^\mu(x) & = & {\cal A}^\mu(x)\,,\nonumber\\
P_-\,A^\mu(x) & = & {{\cal A}^\mu}^*(x)\,.
\eea
With the above results at hand,
we are now in the position 
to give a more concise form of our field-redefinition mapping:
\beq{mapping}
\underline{A}^\mu(x)=e^{-ik \cdot x}\,P_+\,A^\mu(x)+e^{+ik \cdot x}\,P_-\,A^\mu(x)\,.
\eeq
We mention 
that with the field redefinition~\rf{mapping}
and the equations of motion~\rf{lightlikeeom}
one can verify Eq.~\rf{eomA} directly
without using plane-wave expansions. 
We also remark 
that a similar field redefinition exists 
for the closely related Lorentz- and CPT-violating $b^\mu$ parameter 
for SME fermions\cite{rl06}.

\section{Summary and outlook}
\label{conclusions}

The electrodynamics sector of the SME 
contains a Chern--Simons-type operator 
contracted with a Lorentz- and CPT-violating four-vector coupling $k^{\mu}$. 
Such a coupling can, 
for example, 
arise through a nontrivial spacetime topology
or in certain cosmological supergravity models 
as a result of varying scalar fields. 
If $k^\mu$ is lightlike, 
the free solutions of this model 
can be mapped to 
the solutions of conventional Maxwell electrodynamics. 
The specific form of this field redefinition 
is determined by Eq.~\rf{mapping}. 
This mapping is nonlocal, 
and it applies on-shell. 
The existence of such a mapping 
does {\em not} imply that 
the Chern--Simons term is unphysical 
and cannot be measured: 
both off-shell physics 
and interactions 
will typically lead to observable effects.

There are still a few open questions 
regarding this field redefinition 
that need to be addressed in the future. 
It is, 
for instance, 
interesting to determine 
whether this mapping is one-to-one and onto. 
If so, 
there would be a direct correspondence 
between the Lorentz- and CPT-violating Chern--Simons model 
and ordinary Lorentz- and CPT-symmetric electrodynamics. 
Certain known results in the conventional Maxwell model, 
could then simply be translated 
to the more complex Chern--Simons model 
via the field-redefinition map. 
Other open issues concern the question 
as to whether such types of field redefinitions 
can also be found for timelike and spacelike $k^\mu$ 
or for interacting models. 
An additional important aspect 
we have left largely unaddressed 
is gauge symmetry. 
Although we have discussed 
the example of Lorentz gauge, 
a more detailed analysis of 
how gauge conditions are affected by the mapping
could yield valuable insight into the mathematical structure of 
the Chern--Simons model and the field redefinition.

\section*{Acknowledgments}
This work is supported in part 
by the U.S.\ Department of Energy 
under cooperative research agreement No.\ DE-FG02-05ER41360, 
by the European Commission 
under Grant No.\ MOIF-CT-2005-008687, 
by CONACyT under Grant No.\ 55310,  
and by the 
Funda\c{c}\~ao para a Ci\^encia e a Tecnologia
under Grant No.\ CERN/FP/109351/2009.

\end{multicols}
\medline
\begin{multicols}{2}

\end{multicols}

\begin{thebibliography}{99}


\bibitem{ksp} 
See, e.g., 
V.A.\ Kosteleck\'y and S.\ Samuel, 
Phys.\ Rev.\ D {\bf 39}, 683 (1989); 
V.A.\ Kosteleck\'y and R.\ Potting, 
Nucl.\ Phys.\ B {\bf 359}, 545 (1991); 
B.\ Altschul and V.A.\ Kosteleck\'y, 
Phys.\ Lett.\ B {\bf 628}, 106 (2005). 

\bibitem{ncqed} 
See, e.g., 
S.M.\ Carroll \etal, 
Phys.\ Rev.\ Lett.\ {\bf 87}, 141601 (2001); 
Z.\ Guralnik \etal,
Phys.\ Lett.\ B {\bf 517}, 450 (2001); 
C.E.\ Carlson \etal, 
Phys.\ Lett.\ B {\bf 518}, 201 (2001); 
A.\ Anisimov \etal, 
Phys.\ Rev.\ D {\bf 65}, 085032 (2002). 

\bibitem{spacetimevarying} 
V.A.\ Kosteleck\'y \etal, 
Phys.\ Rev.\ D {\bf 68}, 123511 (2003); 
R.\ Jackiw and S.-Y.\ Pi, 
Phys.\ Rev.\ D {\bf 68}, 104012 (2003); 
O.\ Bertolami \etal, 
Phys.\ Rev.\ D {\bf 69}, 083513 (2004); 
N.\ Arkani-Hamed \etal, 
JHEP {\bf 0507}, 029 (2005).

\bibitem{qg}
J.\ Alfaro \etal, 
Phys.\ Rev.\ D {\bf 66}, 124006 (2002); 
D.\ Sudarsky \etal, 
Phys.\ Rev.\ Lett.\ {\bf 89}, 231301 (2002); 
G.\ Amelino-Camelia, 
Mod.\ Phys.\ Lett.\ A {\bf 17}, 899 (2002); 
R.C.\ Myers and M.\ Pospelov, 
Phys.\ Rev.\ Lett.\ {\bf 90}, 211601 (2003); 
N.E.\ Mavromatos, 
Lect.\ Notes Phys.\ {\bf 669}, 245 (2005). 

\bibitem{klink} 
F.R.\ Klinkhamer, 
Nucl.\ Phys.\ B {\bf 578}, 277 (2000); 
F.R.\ Klinkhamer and J.\ Schimmel, 
Nucl.\ Phys.\ B {\bf 639}, 241 (2002). 

\bibitem{fn02} 
C.D.\ Froggatt and H.B.\ Nielsen, 
arXiv:hep-ph/0211106. 

\bibitem{bj} 
J.D.\ Bjorken, 
Phys.\ Rev.\ D {\bf 67}, 043508 (2003). 

\bibitem{brane} 
C.P.\ Burgess \etal, 
JHEP {\bf 0203}, 043 (2002); 
A.R.\ Frey, 
JHEP {\bf 0304}, 012 (2003); 
J.\ Cline and L.\ Valc\'arcel, 
JHEP {\bf 0403}, 032 (2004). 

\bibitem{flatsme}
D.\ Colladay and V.A.\ Kosteleck\'y, 
Phys.\ Rev.\ D {\bf 55}, 6760 (1997);
Phys.\ Rev.\ D {\bf 58}, 116002 (1998); 
V.A.\ Kosteleck\'y and R.\ Lehnert,
Phys.\ Rev.\ D {\bf 63}, 065008 (2001).

\bibitem{curvedsme} 
V.A.\ Kosteleck\'y, 
Phys.\ Rev.\ D {\bf 69}, 105009 (2004); 
R.\ Bluhm and V.A.\ Kosteleck\'y, 
Phys.\ Rev.\ D {\bf 71}, 065008 (2005). 

\bibitem{review}
See, e.g.,  
V.A.\ Kosteleck\'y, ed., 
{\it CPT and Lorentz Symmetry IV}, 
World Scientific, Singapore, 2008; 
D.\ Mattingly, 
Living Rev.\ Rel.\  {\bf 8}, 5 (2005);
V.A.~Kosteleck\'y and N.~Russell,
arXiv:0801.0287v4.

\bibitem{randomphotonexpt}
See, e.g., 
J.\ Lipa \etal,
Phys.\ Rev.\ Lett.\ {\bf 90}, 060403 (2003);
P.L.\ Stanwix \etal,
Phys.\ Rev.\ D {\bf 74}, 081101 (2006);
M.E.\ Tobar \etal,
Phys.\ Rev.\ D {\bf 71}, 025004 (2005);
M.\ Hohensee \etal,
Phys.\ Rev.\ D {\bf 75}, 049902 (2007);
J.P.\ Cotter and B.\ Varcoe,
physics/0603111;
H.\ M\"uller \etal,
Phys.\ Rev.\ Lett.\  {\bf 99}, 050401 (2007);
M.A.~Hohensee \etal,  
Phys.\ Rev.\ Lett.\  {\bf 102}, 170402 (2009);
Phys.\ Rev.\  D {\bf 80}, 036010 (2009);
J.-P.~Bocquet {\it et al.},
Phys.\ Rev.\ Lett.\  {\bf 104}, 241601 (2010).

\bibitem{cherenkov}
See, e.g., 
R.\ Lehnert and R.\ Potting, 
Phys.\ Rev.\ Lett.\ {\bf 93}, 110402 (2004); 
Phys.\ Rev.\ D {\bf 70}, 125010 (2004); 
C.\ Kaufhold and F.R.\ Klinkhamer, 
Nucl.\ Phys.\ B {\bf 734}, 1 (2006); 
B.\ Altschul, 
Phys.\ Rev.\ Lett.\ {\bf 98}, 041603 (2007). 

\bibitem{randomnuexpt}
See, e.g., 
T.\ Katori and R. Tayloe,
in Ref.\ \cite{review};
LSND Collaboration,
L.B.\ Auerbach \etal,
Phys.\ Rev.\ D {\bf 72}, 076004 (2005);
V.A.\ Kosteleck\'y and M.\ Mewes,
Phys.\ Rev.\ D {\bf 69}, 016005 (2004);
V.~Barger \etal,
Phys.\ Lett.\  B {\bf 653}, 267 (2007);
J.S.~D\'{\i}az \etal, 
Phys.\ Rev.\  D {\bf 80}, 076007 (2009).

\bibitem{randomeexpt} 
See, e.g., 
H.\ Dehmelt \etal, 
Phys.\ Rev.\ Lett.\  {\bf 83}, 4694 (1999); 
G.\ Gabrielse \etal, 
Phys.\ Rev.\ Lett.\ {\bf 82}, 3198 (1999); 
L.-S.\ Hou, W.-T.\ Ni, and Y.-C.M.\ Li, 
Phys.\ Rev.\ Lett.\ {\bf 90}, 201101 (2003); 
H.\ M\"uller, 
Phys.\ Rev.\ D {\bf 71}, 045004 (2005). 

\bibitem{randompnexpt} 
See, e.g., 
D.F.\ Phillips \etal, 
Phys.\ Rev.\ D {\bf 63}, 111101(R) (2001); 
P.\ Wolf \etal, 
Phys.\ Rev.\ Lett.\ {\bf 96}, 060801 (2006); 
O.\ Bertolami \etal, 
Phys.\ Lett.\ B {\bf 395}, 178 (1997); 
I.~Altarev {\it et al.},
Phys.\ Rev.\ Lett.\  {\bf 103}, 081602 (2009);
Europhys.\ Lett.\  {\bf 92}, 51001 (2010).

\bibitem{randomhadronexpt} 
See, e.g., 
BELLE Collaboration, 
K.\ Abe \etal, 
Phys.\ Rev.\ Lett.\ {\bf 86}, 3228 (2001); 
BaBar Collaboration, 
B.\ Aubert \etal, 
Phys.\ Rev.\ Lett.\ {\bf 92}, 181801 (2004); 
arXiv:hep-ex/0607103; 
FOCUS Collaboration, 
J.M.\ Link \etal, 
Phys.\ Lett.\ B {\bf 556}, 7 (2003); 
G.~Amelino-Camelia {\it et al.},
Eur.\ Phys.\ J.\  C {\bf 68}, 619 (2010).

\bibitem{muexpt} 
V.W.\ Hughes \etal, 
Phys.\ Rev.\ Lett.\ {\bf 87}, 111804 (2001); 
R.\ Bluhm \etal, 
Phys.\ Rev.\ Lett.\ {\bf 84}, 1098 (2000). 

\bibitem{gravity}
V.A.~Kosteleck\'y \etal,
Phys.\ Rev.\ Lett.\  {\bf 100}, 111102 (2008);
V.A.~Kosteleck\'y and J.~Tasson, 
Phys.\ Rev.\ Lett.\  {\bf 102}, 010402 (2009);
Phys.\ Rev.\  D {\bf 83}, 016013 (2011);
J.B.R.~Battat \etal, 
Phys.\ Rev.\ Lett.\  {\bf 99}, 241103 (2007);
Q.G.~Bailey,
Phys.\ Rev.\  D {\bf 80}, 044004 (2009).

\bibitem{investigations}
See, e.g., 
R.~Jackiw and V.A.~Kosteleck\'y,
Phys.\ Rev.\ Lett.\  {\bf 82}, 3572 (1999);
V.A.\ Kosteleck\'y \etal, 
Phys.\ Rev.\ D {\bf 65}, 056006 (2002); 
R.\ Lehnert, 
Phys.\ Rev.\ D {\bf 68}, 085003 (2003); 
J.\ Math.\ Phys.\ {\bf 45}, 3399 (2004); 
V.A.\ Kosteleck\'y and R.\ Potting, 
Gen.\ Rel.\ Grav.\  {\bf 37}, 1675 (2005); 
B.~Altschul,
Phys.\ Rev.\ D {\bf 73}, 045004 (2006);
J.\ Phys.\ A {\bf 39}, 13757 (2006);
J.~Alfaro \etal,  
Phys.\ Lett.\  B {\bf 639}, 586 (2006);
Int.\ J.\ Mod.\ Phys.\  A {\bf 25}, 3271 (2010);
J.~Alfaro and L.F.~Urrutia,
Phys.\ Rev.\  D {\bf 81}, 025007 (2010);
Q.G.~Bailey, 
Phys.\ Rev.\  D {\bf 82}, 065012 (2010).

\bibitem{rl06}
R.\ Lehnert, 
Phys.\ Rev.\  D {\bf 74}, 125001 (2006).

\bibitem{hariton07}
A.J.~Hariton and R.~Lehnert,
Phys.\ Lett.\  A {\bf 367}, 11 (2007).

\bibitem{susy} 
M.S.\ Berger and V.A.\ Kosteleck\'y, 
Phys.\ Rev.\ D {\bf 65}, 091701(R) (2002); 
H.\ Belich \etal, 
Phys.\ Lett.\ A {\bf 370}, 126 (2007);
M.S.\ Berger,
Phys.\ Rev.\ D {\bf 68}, 115005 (2003). 

\bibitem{km02} 
V.A.\ Kosteleck\'y and M.\ Mewes, 
Phys.\ Rev.\ D {\bf 66}, 056005 (2002). 

\bibitem{cg99} 
S.\ Coleman and S.L.\ Glashow, 
Phys.\ Rev.\ D {\bf 59}, 116008 (1999).

\bibitem{mcs}
S.M.~Carroll \etal,
Phys.\ Rev.\  D {\bf 41}, 1231 (1990).

\bibitem{adamklink01}
C.~Adam and F.R.~Klinkhamer,
Nucl.\ Phys.\  B {\bf 607}, 247 (2001).

\end{thebibliography}
\end{document}